\documentclass{ws-p8-50x6-00}

\begin{document}

\title{Quantum Energies of Solitons}

\author{Herbert Weigel\thanks{Heisenberg--Fellow}}

\address{Institute for Theoretical Physics,~T\"ubingen University\\
Auf der Morgenstelle 14,~D--72076 T\"ubingen,~Germany\\
E-mail: herbert.weigel@uni-tuebingen.de}

\maketitle

\abstracts{
For renormalizable models a method is presented to unambiguously
compute the energy that is carried by localized field configurations
(solitons). A variational approach for the total energy is utilized to 
search for soliton configurations. As an example a $1+1$ dimensional 
model is considered. The quantum energy of configurations that are 
translationally invariant for a subset of coordinates is discussed.}

\noindent
In this talk I present results that emerged from the 
collaboration\cite{MIT1,MIT2,MIT3} with E.~Farhi, N. Graham, 
R. L. Jaffe and M. Quandt.

\section{Introduction}

Field theories can contain spatially varying (but time independent)
configurations $\Phi$ that are local minima of the classical energy. 
When quantum effects are taken into account, the classical 
description must be re--examined because the spatially varying soliton 
configuration should minimize the total energy
\begin{equation}
E_{\rm tot}[\Phi]=E_{\rm cl}[\Phi]+E_{\rm qu}[\Phi]
\label{etot}
\end{equation}
which takes into account classical ($E_{\rm cl}$) and quantum 
($E_{\rm qu}$) contributions. Since the total energy for general
configurations is difficult to compute, quantum effects are typically
computed as approximate corrections to the classical soliton. In this
talk I describe an approach that unambiguously yields the total
energy up to one loop order, {\it i.e.} ${\cal O}(\hbar)$.

At ${\cal O}(\hbar)$ the quantum
contribution is the sum of the change of the vacuum energy and the 
(local) counterterm functional 
\begin{equation}
E_{\rm qu}[\Phi]=E_{\rm vac}[\Phi]+E_{\rm c.t.}[\Phi]\, .
\label{eqf}
\end{equation}
Formally $E_{\rm vac}$ is given as the sum of the changes of the 
frequencies $\omega$ of the small amplitude fluctuations $\eta$ about 
$\Phi$. These frequencies are determined from a Schr\"odinger--type 
wave--equation 
\begin{equation}
\left\{-\vec{\partial}\,^2+m^2+U(\Phi)\right\}\eta=\omega^2\eta\, ,
\label{schrodinger}
\end{equation}
with a potential $U$ that is obtained from $\Phi$. The evaluation of 
$E_{\rm vac}$ is \underline{non--perturbative} as can {\it e.g.}~be 
observed from the appearance of bound states in Eq.~(\ref{schrodinger}). 
$E_{\rm vac}$ is ultra--violet divergent and requires 
regularization. On the other hand, $E_{\rm c.t.}$ is computed as a local 
integral of $\Phi$ (and derivatives thereof) that contains 
divergent coefficients. Commonly these coefficients are determined in 
the \underline{perturbative} 
sector (where $\Phi$ is set to its vacuum value) using 
{\it e.g.} dimensional regularization. The main problem is to find 
a regularization scheme for $E_{\rm vac}$ that is compatible with
the determination of the counterterms such that the sum~(\ref{eqf})
is finite and unique for any prescribed background $\Phi$. 

\section{The Phase Shift Approach}

The vacuum energy acquires contributions from states that are bound 
in the potential $U(\Phi)$ as well as from scattering states. While the 
former contribution can be expressed as
a finite sum over discrete levels the latter may be computed from
the change of the density of states. This change is given in terms
of the derivative of the phases shifts $\delta(k)$. Thus I may formally 
write\footnote{For a fermion loop an overall sign emerges.}
\begin{equation}
E_{\rm vac}\, \sim\, \frac{1}{2}\sum_i^{\rm b.s.}|\omega_i|
+\int_0^\infty\frac{dk}{2\pi}\omega_k\, \frac{d}{dk}
\sum_\ell D_\ell\delta_\ell(k)\, ,
\label{evacform}
\end{equation}
where $\omega_i$ are the eigen--frequencies of the bound states
and $D_\ell$ is the degeneracy factor associated with the channels
labeled by $\ell$ (e.g. $D_\ell=2\ell+1$ if $\ell$ refers to 
orbital angular momentum). Furthermore $\omega_k=\sqrt{k^2+m^2}$.
As already noted, the above integral is ultra--violet divergent. 
Fortunately the large $k$ behavior of the phase shifts can be
isolated using the Born--series. This series represents an 
expansion of the phase shifts in terms of the potential $U$. In 
general this expansion does \underline{not} converge for all
$k$, however, it does for large enough $k$. The expansion
of the vacuum energy in terms of $U$ can also be obtained 
from the expansion of 
${\rm ln}\,{\rm det}(-\vec{\partial}\,^2+m^2+U)$ that corresponds
to the evaluation of a set of Feynman diagrams. The identity of these
expansions has been established utilizing dimensional 
regularization\cite{MIT1,MIT2}. Now the central 
idea\cite{Fa98,Schw_Baa,MIT1} is to add and subtract the 
expansion in $U=U(\Phi)$. This yields 
\begin{eqnarray}
E_{\rm qu}&=&
\sum_i^{\rm b.s.}\left|\frac{\omega_i}{2}\right|
+\int_0^\infty\frac{dk}{2\pi}\,\omega_k\, \frac{d}{dk}
\sum_\ell D_\ell\left\{\delta_\ell(k)-\delta^{(1)}_\ell(k)
-\delta^{(2)}_\ell(k)-\ldots\right\}
\nonumber \\* && \hspace{2.5cm}
+\quad \begin{minipage}{5cm}
\epsfig{file=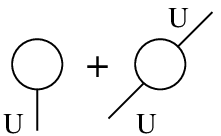}
\end{minipage}\hspace{-3cm}+\ldots\, +E_{\rm c.t.} ,
\label{eqf1}
\end{eqnarray}
where $U=U(\Phi)$ denotes the potential generated by the background field.
The divergent Feynman diagrams in Eq.~(\ref{eqf1}) can be computed using 
standard techniques and, when  added to the counterterm 
contribution $E_{\rm c.t.}$, a finite and unique result for the total 
energy is obtained. 

\section{Chiral Model in D=1+1 as an Example}

Now I would like to apply this formalism to a simple chiral
model in $D=1+1$. In this model a two--component boson 
field $\vec{\phi}=(\phi_1,\phi_2)$ couples chirally to a 
fermion $\Psi$ that come in $N_f$ (equivalent) modes:
\begin{eqnarray}
{\cal L}=\frac{1}{2}\, \partial_\mu\vec{\phi}\cdot\partial^\mu\vec{\phi}
+\sum_{n=1}^{N_f}\bar{\Psi}_i\left\{i\partial \hskip -0.5em /
- G\left(\phi_1+i\gamma_5\phi_2\right) \right\}\Psi_i\, .
\label{lagd11}
\end{eqnarray}
where the potential for the boson field
\begin{eqnarray}
V(\vec{\phi})=\frac{\lambda}{8}
\left[\vec{\phi}\cdot\vec{\phi}
-v^2+\frac{2\alpha v^2}{\lambda}\right]^2
-\alpha v^3\left(\phi_1-v\right)+{\rm const.}\,
\label{potential}
\end{eqnarray}
contains a term (proportional to $\alpha$) that breaks the chiral symmetry 
explicitly in order to avoid problems stemming from (unphysical) infra--red 
singularities that occur when the vacuum configuration would be determined 
via the na{\"\i}ve treatment of spontaneous symmetry breaking\cite{Col73}.
In this manner it is guaranteed that the VEV is given by
$\langle \vec{\phi}\rangle=(v,0)$. Here the counterterm Lagrangian is
not presented explicitly. It is determined such that the quantum 
corrections lead to a vanishing tadpole diagram for the boson field.
Note that considering only the classical contribution 
does \underline{not} support a stable soliton soliton.

In the limit that the number of fermion modes becomes large with 
$v^2/N_f\sim{\cal O}(1)$ only the classical and one fermion loop pieces 
contribute. In the following I will only consider that limit, 
{\it i.e.} $E_{\rm tot}=E_{\rm cl}+E_{\rm F}$. The fermion contribution 
can be split into two pieces $E_{\rm F}=E_{\rm vac}+E_{\rm val}$. 
The valence part $E_{\rm val}$ is given in terms
of the bound state energies such as to saturate the total fermion number
that is fixed to be $N_F$. The vacuum piece is computed according to
the formalism described in the preceding section:
\begin{eqnarray}
E_{\rm vac}=-\frac{1}{2}\sum_i^{\rm b.s.}\left(
\left|\omega_i\right|-Gv\right)
-\int_0^\infty\frac{dk}{2\pi}\, \left(\omega_k-Gv\right)
\frac{d}{dk}\left(\delta_{\rm F}(k)-\delta^{(1)}(k)\right)\, ,
\label{efermion}
\end{eqnarray}
which is obtained from Eq~(\ref{eqf1}) by employing Levinson's theorem.
Here $\delta_{\rm F}$ denotes the sum of the eigenphase shifts\footnote{The
eigen--channels are labeled by parity and the sign of the single particle 
eigen--energies.}. The subtraction 
\begin{equation}
\delta^{(1)}(k)=\frac{2G^2}{k}\int_0^\infty dx 
\left(v^2-\vec{\phi}\,^2(x)\right)
\label{subtr}
\end{equation}
that renders $E_{\rm vac}$ finite
contains both first and second order Born approximants in the fluctuations
of $\vec{\phi}$ about $\langle \vec{\phi}\rangle$. The first order 
is unambiguously fixed by the no--tadpole renormalization condition and
the second order by the chiral symmetry. 

\begin{figure}[ht]
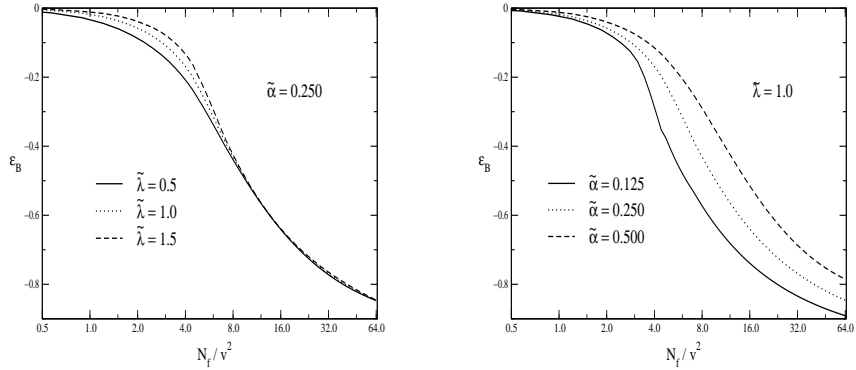

\centerline{
\epsfig{figure=lambda.eps,height=5.0cm,width=5.5cm,angle=270}
\hspace{1cm}
\epsfig{figure=alpha.eps,height=5.0cm,width=5.5cm,angle=270}
}
\caption{\label{fig_1}The maximal binding energy as a function of the model 
parameters as obtained from the {\it Ansatz}~(\protect\ref{variation}) 
in units of $Gv$; $\tilde{\alpha}={\alpha}/{G^2}$ and
$\tilde{\lambda}={\lambda}/{G^2}$.}
\end{figure}

Having established the energy functional I now consider
variational {\it Ans\"atze} for the background field that turn this
functional in a function of the variational parameters. As an example 
I assume
\begin{equation}
\phi_1+i\phi_2=v\left\{1-R+R\,{\rm exp}\hspace{0.1mm}
\left[i\pi\left(1+{\rm tanh}(Gvx/w)\right)\right]\right\}
\label{variation}
\end{equation}
that introduces width ($w$) and amplitude ($R$) parameters. For
prescribed model parameters ($G$,$v$,etc.) the energy must be
minimized with respect to $w$ and $R$. The resulting binding 
energy $\epsilon_{\rm B}=E_{\rm tot}-Gv$ is shown in figure~\ref{fig_1}.
Even though the {\it Ansatz}~(\ref{variation}) may not be the final answer
to the minimalization problem, $\epsilon_{\rm B}$ is definitely negative.
Thus a solitonic configuration is energetically favored showing that 
indeed quantum fluctuations can create a soliton that is not stable at
the classical level.

\section{Quantum Energies of Interfaces}

In some cases ({\it e.g.}~domain walls) the background potential in 
the Schr\"odinger--type equation depends only on a subset of the 
coordinates $U(x,\vec{y})=u(x)$, where the number of ``trivial''
dimensions is $n$: $\vec{y}=(y_1,\ldots,y_n)$. I will refer to such 
configurations as interfaces. The single particle energies 
then parametrically depend
on the momentum $\vec{p}$ conjugate to $\vec{y}$:
$\omega(p,k)=\sqrt{\vec{p}\,^2+m^2+k^2}$ and
$\omega_i(p)=\sqrt{\vec{p}\,^2+m^2-\kappa_i^2}$,
where $k$ and $\kappa_i$ 
label the scattering and bound states obtained from the one--dimensional
Schr\"odinger--type equation for $u(x)$. However,
the phase shifts depend only on the momentum $k$ conjugate to $x$:
$\delta=\delta(k)$. When straightforwardly integrating the $n+1$
dimensional generalization of eq~(\ref{eqf1}) over $\vec{p}$
a logarithmic singularity seems to emerge at $n=1$ because
$\omega(p,k)-\sqrt{\vec{p}\,^2+m^2}\to k^2/2|\vec{p}|$ for large 
$|\vec{p}|$. This cannot be regularized by Born--subtractions that 
only involve $k$. As 
in dimensional regularization it is suitable to consider $n<1$. Other 
cases are obtained by analytic continuation. This yields the quantum 
energy (density) of the interface:
\begin{eqnarray}
\epsilon(n)&=&-\frac{\Gamma\left(-\frac{n+1}{2}\right)}
{(4\pi)^{\frac{n+1}{2}}}\left\{\int_0^\infty \frac{dk}{\pi}
\left(\omega_k^{n+1}-m^{n+1}\right)\frac{d}{dk} \bar{\delta}(k)
+\sum_i^{\rm b.s.}\left(\omega_i^{n+1}-m^{n+1}\right)\right\}
\nonumber \\ &&\hspace{2cm}
+\epsilon_{\rm F.D.}+\epsilon_{\rm c.t.}
\label{interf1}
\end{eqnarray}
where $\bar{\delta}(k)$ denote the Born--subtracted phase shift 
and $\epsilon_{\rm F.D.}$ is the Feynman diagram contribution.
While the known formulae, {\it e.g.} Eq~(\ref{efermion}) are recovered 
for $n=0$, the singularity for $n\to1$ is reflected as a pole in the 
$\Gamma$ function. Consistency conditions\footnote{For example, 
a $\phi^4$--theory in $D=2+1$ must be renormalizable.} require 
$\epsilon(1)$ to be finite. Hence the residuum of the pole in the 
$\Gamma$--function must vanish, {\it i.e.}
\begin{equation}
\int_0^\infty\frac{dk}{\pi}k^2\frac{d}{dk}\bar{\delta}(k)
-\sum_i^{\rm b.s.}\kappa_i^2=0\, .
\label{res}
\end{equation}
In this way a number of sum rules\footnote{They can also be proven
with Jost--function techniques\cite{Puff75,MIT3}.} between bound state 
energies $\sqrt{m^2-\kappa_i^2}$ and (Born--subtracted) phase shifts 
can be established. Ultimately the limit $n\to1$ can safely be assumed
yielding the interface energy
\begin{eqnarray}
\lim_{n\to1}\epsilon(n)&=&-\frac{1}{4\pi} \left\{
\int_0^\infty \frac{dk}{\pi} \left(\omega(k)^2 \log
\frac{\omega(k)^2}{4 \pi \mu^2} - m^2 \log \frac{m^2}{4 \pi \mu^2}\right)
\frac{d}{dk} \bar \delta(k) \right. 
\hspace{1cm}\nonumber \\* && \hspace{1.0cm} \left.
+ \sum_i \left(\omega_i^2\log\frac{\omega_i^2}{4 \pi \mu^2} -
m^2\log \frac{m^2}{4 \pi \mu^2} \right) \right\}
+\epsilon_{\rm F.D.}+\epsilon_{\rm c.t.}
\label{master}
\end{eqnarray}
The scale $\mu$ has been introduced for dimensional reasons. It is
arbitrary because its contribution is proportional to the 
residuum~(\ref{res}).

\section{Summary}

In this talk I have presented a formalism to unambiguously and
numerically feasibly compute the one--loop quantum corrections
to energies of spatially varying field configurations. This approach 
strongly relies on the identity of Feynman diagrams and Born approximants 
to Casimir energies. Utilizing a variational approach to the total 
energy solitons can be constructed. As an example I 
have shown that in a $1+1$ dimensional chiral model quantum 
corrections create a soliton that is classically unstable.
Finally I have derived a master formula for interface energies
whose consistency demands sum rules for scattering data that can
be considered generalizations of Levinson's Theorem.

\section*{Acknowledgments}
I would like to thank the organizers of the workshop for the pleasant
and stimulating working atmosphere. In particular I am grateful to 
B. M\"uller for bringing Ref.\cite{Puff75} to my attention.
Furthermore I would like to thank E. Farhi, N. Graham, R. L. Jaffe 
and M. Quandt for the fruitful collaboration. This work is supported in 
part by the Deutsche Forschungsgemeinschaft under contracts We 1254/3-1,4-2.

\end{document}